\documentclass{emulateapj}

\submitted{Accepted in ApJ}
\usepackage{graphicx}        
\usepackage{bm}              
\usepackage{natbib}
\usepackage{lscape}
\usepackage[mediumspace,Gray,squaren]{SIunits}

\shorttitle{SZ-Selected Galaxy Clusters from ACT}
\shortauthors{Marriage et al.}

\newcommand{\arone}{148\,GHz}
\newcommand{\artwo}{218\,GHz}
\newcommand{\arthree}{277\,GHz}

\newcommand{\commentx}[1]{}

\renewcommand{\vec}[1]{\mbox{\boldmath$#1$}} 

\newcommand{\ra}[3]   
   {\makebox[1.5em][r]{#1}\makebox[1.5em][r]{#2} \makebox[2em][r]{#3}}

\newcommand{\hms}[3]  
   {${#1}^{\mathrm{h}}{#2}^{\mathrm{m}}{#3}^{\mathrm{s}}$}

\newcommand{\hmin}[2]  
   {\ensuremath{{#1}^{\mathrm{h}}{#2}^{\mathrm{m}}}}
   
\newcommand{\hours}[1]  
   {\ensuremath{{#1}^{\mathrm{h}}}}

\newcommand{\dms}[3]  
   {\ensuremath{{#1}\degree{#2}\arcminute{#3}\arcsecond}}

\newcommand{\dm}[2]  
   {\ensuremath{{#1}\degree{#2}\arcminute}}

\newcommand{\ukcmb}  
           {\ensuremath{\micro \kelvin_\mathrm{cmb}}}

\newcommand{\uk}  
           {\ensuremath{\micro \kelvin}}

\newcommand{\fdeg} 
           {\hbox{$.\!\!^{\circ}$}}

\usepackage{color}

\newcommand{\numberOfClusters}{twenty-three}
\newcommand{\NumberOfNewClusters}{Ten}

\newcommand{\highSNR}{6}
\newcommand{\numberOfHighSNRClusters}{nine}

\newcommand{\numberOfNewHighSNRClusters}{three}

\newcommand{\numberOfLowSNRClusters}{fourteen}

\newcommand{\numberOfNewLowSNRClusters}{seven}

\newcommand{\redshiftRange}{$z>0.1$}
\newcommand{\minimumMass}{$2 \times 10^{14}\,M_\odot$}
\newcommand{\associationRadius}{$3\arcmin$}
\newcommand{\highSNRPurity}{95\%}
\newcommand{\completenessRange}{$M_{500c} > 6\times10^{14} M_\odot$}
\newcommand{\highSNRCompleteness}{$\sim$45\%}
\newcommand{\totalSampleCompleteness}{$\sim$80\%}

\newcommand{\medianNoiseOneArcminute}{$36 \micro{\rm K}$}

\begin{document}

\title{The Atacama Cosmology Telescope: Sunyaev-Zel'dovich Selected Galaxy Clusters at \arone\ in the 2008 Survey}

\author{
Tobias~A.~Marriage\altaffilmark{1,2},
Viviana~Acquaviva\altaffilmark{3,1},
Peter~A.~R.~Ade\altaffilmark{4},
Paula~Aguirre\altaffilmark{5},
Mandana~Amiri\altaffilmark{6},
John~William~Appel\altaffilmark{7},
L.~Felipe~Barrientos\altaffilmark{5},
Elia~S.~Battistelli\altaffilmark{8,6},
J~Richard~Bond\altaffilmark{9},
Ben~Brown\altaffilmark{10},
Bryce~Burger\altaffilmark{6},
Jay~Chervenak\altaffilmark{11},
Sudeep~Das\altaffilmark{12,7,1},
Mark~J.~Devlin\altaffilmark{13},
Simon~R.~Dicker\altaffilmark{13},
W.~Bertrand~Doriese\altaffilmark{14},
Joanna~Dunkley\altaffilmark{15,7,1},
Rolando~D\"{u}nner\altaffilmark{5},
Thomas~Essinger-Hileman\altaffilmark{7},
Ryan~P.~Fisher\altaffilmark{7},
Joseph~W.~Fowler\altaffilmark{14,7},
Amir~Hajian\altaffilmark{9,1,7},
Mark~Halpern\altaffilmark{6},
Matthew~Hasselfield\altaffilmark{6},
Carlos~Hern\'andez-Monteagudo\altaffilmark{16},
Gene~C.~Hilton\altaffilmark{14},
Matt~Hilton\altaffilmark{17,18},
Adam~D.~Hincks\altaffilmark{7},
Ren\'ee~Hlozek\altaffilmark{15},
Kevin~M.~Huffenberger\altaffilmark{19},
David~Handel~Hughes\altaffilmark{20},
John~P.~Hughes\altaffilmark{3},
Leopoldo~Infante\altaffilmark{5},
Kent~D.~Irwin\altaffilmark{14},
Jean~Baptiste~Juin\altaffilmark{5},
Madhuri~Kaul\altaffilmark{13},
Jeff~Klein\altaffilmark{13},
Arthur~Kosowsky\altaffilmark{10},
Judy~M~Lau\altaffilmark{21,22,7},
Michele~Limon\altaffilmark{23,13,7},
Yen-Ting~Lin\altaffilmark{24,1,5},
Robert~H.~Lupton\altaffilmark{1},
Danica~Marsden\altaffilmark{13},
Krista~Martocci\altaffilmark{25,7},
Phil~Mauskopf\altaffilmark{4},
Felipe~Menanteau\altaffilmark{3},
Kavilan~Moodley\altaffilmark{17,18},
Harvey~Moseley\altaffilmark{11},
Calvin~B~Netterfield\altaffilmark{26},
Michael~D.~Niemack\altaffilmark{14,7},
Michael~R.~Nolta\altaffilmark{9},
Lyman~A.~Page\altaffilmark{7},
Lucas~Parker\altaffilmark{7},
Bruce~Partridge\altaffilmark{27},
Hernan~Quintana\altaffilmark{5},
Erik~D.~Reese\altaffilmark{13},
Beth~Reid\altaffilmark{28,7},
Neelima~Sehgal\altaffilmark{21},
Blake~D.~Sherwin\altaffilmark{7},
Jon~Sievers\altaffilmark{9},
David~N.~Spergel\altaffilmark{1},
Suzanne~T.~Staggs\altaffilmark{7},
Daniel~S.~Swetz\altaffilmark{13,14},
Eric~R.~Switzer\altaffilmark{25,7},
Robert~Thornton\altaffilmark{13,29},
Hy~Trac\altaffilmark{30,31},
Carole~Tucker\altaffilmark{4},
Ryan~Warne\altaffilmark{17},
Grant~Wilson\altaffilmark{32},
Ed~Wollack\altaffilmark{11},
Yue~Zhao\altaffilmark{7}
}
\altaffiltext{1}{Department of Astrophysical Sciences, Peyton Hall, 
Princeton University, Princeton, NJ USA 08544}
\altaffiltext{2}{Current address: Dept. of Physics and Astronomy, Johns Hopkins University, 3400 N. Charles St., Baltimore, MD 21218-2686}
\altaffiltext{3}{Department of Physics and Astronomy, Rutgers, 
The State University of New Jersey, Piscataway, NJ USA 08854-8019}
\altaffiltext{4}{School of Physics and Astronomy, Cardiff University, The Parade, 
Cardiff, Wales, UK CF24 3AA}
\altaffiltext{5}{Departamento de Astronom{\'{i}}a y Astrof{\'{i}}sica, 
Facultad de F{\'{i}}sica, Pontific\'{i}a Universidad Cat\'{o}lica,
Casilla 306, Santiago 22, Chile}
\altaffiltext{6}{Department of Physics and Astronomy, University of
British Columbia, Vancouver, BC, Canada V6T 1Z4}
\altaffiltext{7}{Joseph Henry Laboratories of Physics, Jadwin Hall,
Princeton University, Princeton, NJ, USA 08544}
\altaffiltext{8}{Department of Physics, University of Rome ``La Sapienza'', 
Piazzale Aldo Moro 5, I-00185 Rome, Italy}
\altaffiltext{9}{Canadian Institute for Theoretical Astrophysics, University of
Toronto, Toronto, ON, Canada M5S 3H8}
\altaffiltext{10}{Department of Physics and Astronomy, University of Pittsburgh, 
Pittsburgh, PA, USA 15260}
\altaffiltext{11}{Code 553/665, NASA/Goddard Space Flight Center,
Greenbelt, MD, USA 20771}
\altaffiltext{12}{Berkeley Center for Cosmological Physics, LBL and
Department of Physics, University of California, Berkeley, CA, USA 94720}
\altaffiltext{13}{Department of Physics and Astronomy, University of
Pennsylvania, 209 South 33rd Street, Philadelphia, PA, USA 19104}
\altaffiltext{14}{NIST Quantum Devices Group, 325
Broadway Mailcode 817.03, Boulder, CO, USA 80305}
\altaffiltext{15}{Department of Astrophysics, Oxford University, Oxford, 
UK OX1 3RH}
\altaffiltext{16}{Max Planck Institut f\"ur Astrophysik, Postfach 1317, 
D-85741 Garching bei M\"unchen, Germany}
\altaffiltext{17}{Astrophysics and Cosmology Research Unit, School of
Mathematical Sciences, University of KwaZulu-Natal, Durban, 4041,
South Africa}
\altaffiltext{18}{Centre for High Performance Computing, CSIR Campus, 15 Lower
Hope St., Rosebank, Cape Town, South Africa}
\altaffiltext{19}{Department of Physics, University of Miami, Coral Gables, 
FL, USA 33124}
\altaffiltext{20}{Instituto Nacional de Astrof\'isica, \'Optica y 
Electr\'onica (INAOE), Tonantzintla, Puebla, Mexico}
\altaffiltext{21}{Kavli Institute for Particle Astrophysics and Cosmology, Stanford
University, Stanford, CA, USA 94305-4085}
\altaffiltext{22}{Department of Physics, Stanford University, Stanford, CA, 
USA 94305-4085}
\altaffiltext{23}{Columbia Astrophysics Laboratory, 550 W. 120th St. Mail Code 5247,
New York, NY USA 10027}
\altaffiltext{24}{Institute for the Physics and Mathematics of the Universe, 
The University of Tokyo, Kashiwa, Chiba 277-8568, Japan}
\altaffiltext{25}{Kavli Institute for Cosmological Physics, 
5620 South Ellis Ave., Chicago, IL, USA 60637}
\altaffiltext{26}{Department of Physics, University of Toronto, 
60 St. George Street, Toronto, ON, Canada M5S 1A7}
\altaffiltext{27}{Department of Physics and Astronomy, Haverford College,
Haverford, PA, USA 19041}
\altaffiltext{28}{Institut de Ciencies del Cosmos (ICC), University of
Barcelona, Barcelona 08028, Spain}
\altaffiltext{29}{Department of Physics , West Chester University 
of Pennsylvania, West Chester, PA, USA 19383}
\altaffiltext{30}{Department of Physics, Carnegie Mellon University, Pittsburgh, PA 15213}
\altaffiltext{31}{Harvard-Smithsonian Center for Astrophysics, 
Harvard University, Cambridge, MA, USA 02138}
\altaffiltext{32}{Department of Astronomy, University of Massachusetts, 
Amherst, MA, USA 01003}

\begin{abstract}
We report on \numberOfClusters\ clusters detected blindly as Sunyaev-Zel'dovich (SZ) decrements in a 148\,GHz, 455 square-degree map of the southern sky made with data from the Atacama Cosmology Telescope 2008 observing season. All SZ detections announced in this work have confirmed optical counterparts. \NumberOfNewClusters\ of the clusters are new discoveries. One newly discovered cluster, ACT-CL J0102-4915, with a redshift of 0.75 (photometric), has an SZ decrement comparable to the most massive systems at lower redshifts. Simulations of the cluster recovery method reproduce the sample purity measured by optical follow-up. In particular, for clusters detected with a signal-to-noise ratio greater than six, simulations are consistent with optical follow-up that demonstrated this  subsample is 100\% pure. The simulations further imply that the total sample is 80\% complete for clusters with mass in excess of $6 \times 10^{14}$ solar masses referenced to the cluster volume characterized by five hundred times the critical density. The Compton $y$ -- X-ray luminosity mass comparison for the eleven best detected clusters visually agrees with both self-similar and non-adiabatic, simulation-derived scaling laws.
\end{abstract}


\maketitle


\section{INTRODUCTION}

\setcounter{footnote}{0}

Measuring the redshift evolution of the mass function of galaxy clusters 
provides powerful constraints on cosmological parameters, complementary to 
those obtained from different measurements, such as the angular power 
spectrum of the cosmic microwave background \citep[e.g.,][]{komatsu/etal:prep} or luminosity distances of Type Ia supernovae \citep[e.g.,][]{hicken/etal:2009}. The 
potential of cluster surveys as a cosmological probe has recently been 
demonstrated through the analysis of large X-ray and optically selected 
cluster samples spanning a wide range in redshift \citep[e.g.,][]{vikhlinin/etal:2009, mantz/etal:2010, rozo/etal:2010}.

\begin{figure*}[ht]
\begin{center}
  \resizebox{\textwidth}{!}{
    \plotone{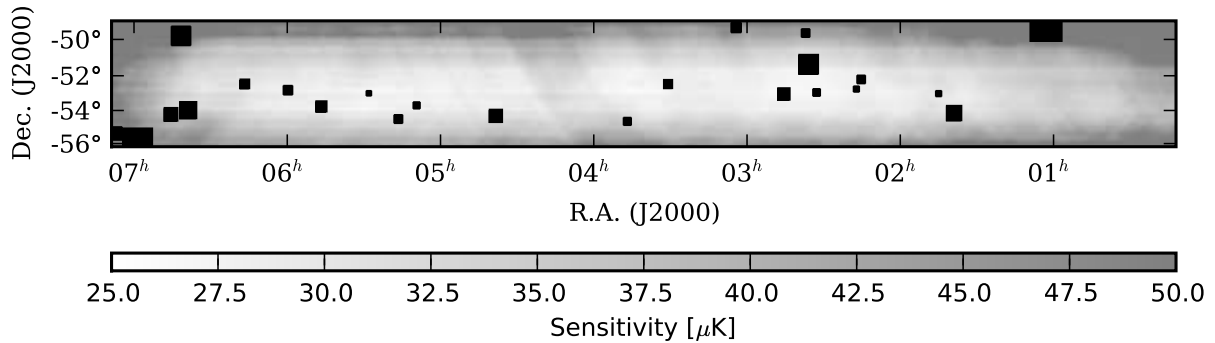}
  }
\caption{Sensitivity map with detections. The map shows the sensitivity over the subset of ACT 2008 \arone\ dataset  considered for this study. The gray-scale encodes the noise rms in $\micro{\rm K}$ of a map match-filtered for a $\beta$-model ($\beta=0.86$, $\theta_{\rm c}=1.00\arcmin$). The median noise in the map is \medianNoiseOneArcminute. Black boxes mark the location of the \numberOfClusters\ optically confirmed clusters. The size of each box is proportional to the corresponding cluster decrement. }
\label{fig:coverage}
\end{center}
\end{figure*}

A powerful method for cluster detection and characterization is provided by the thermal Sunyaev-Zel'dovich (SZ) effect. The SZ effect is the process by which energy is transferred from the hot intra-cluster medium (ICM) of galaxy clusters to the cosmic microwave background (CMB) through inverse Compton scattering \citep{zeldovich/sunyaev:1969,sunyaev/zeldovich:1970}. The overall effect is to distort the thermal spectrum of the CMB. In the case of the present study at \arone, the SZ distortion produces a roughly arcminute-scale temperature decrement along the line of sight to a galaxy cluster. The SZ distortion is proportional to the pressure of the ICM integrated along the line of sight, providing a unique probe of cluster physics.  The integrated SZ flux therefore scales as the thermal energy content of the cluster, which is expected to scale with cluster mass for cluster gas hydrostatically supported by thermal pressure. The amplitude of the SZ distortion is approximately independent of the cluster redshift, in marked contrast to the observed cluster brightness in other wave bands. Therefore, the SZ effect is particularly suited for inventorying the
distant universe.  Galaxy cluster surveys are one of the few methods
that probe the growth of structure and interpretation of survey yields
has the potential to precisely constrain the equation of state of dark energy \citep[e.g.,][]{bartlett/silk:1994, holder/etal:2000, haiman/mohr/holder:2001,
majumdar/mohr:2004}. For a full description of the SZ effect, historical perspectives, and applications to cosmology, see the review articles by \cite{bernstein/dodelson:1990}, \cite{birkinshaw:1991}, \cite{rephaeli:1995}, \cite{birkinshaw:1999}, and \cite{carlstrom/holder/reese:2002}.

The history of successful targeted observations of the SZ effect in previously known galaxy clusters goes back two decades. Pioneering measurements were made with the OVRO 40m telescope \citep{birkinshaw/hughes/arnaud:1991}, the Ryle Telescope \citep{ jones/etal:1993}, the OVRO 5m telescope \citep{ herbig/etal:1995, myers/etal:1997}, the OVRO/BIMA interferometers \citep{carlstrom/joy/grego:1996}, the bolometric SUZIE receiver on the Caltech Submillimeter Observatory \citep{holzapfel/etal:1997}, and the Nobeyama Radio Observatory 45\,m telescope \citep{komatsu/etal:1999}. Scientifically these original studies put special emphasis on deriving the Hubble parameter from angular diameter distances to clusters measured through a combination of SZ and X-ray data. In the last decade, these geometric measurements have been expanded to statistically significant samples of clusters (e.g., the OVRO/BIMA studies by \citealt{reese/etal:2002} and \citealt{bonamente/etal:2006}) and serve as an independent check on estimates that rely on the conventional extragalactic distance ladder (e.g., \citealt{ freedman/etal:2001, riess/etal:2009}). Beyond distance scale studies, targeted SZ work has focused on understanding the SZ correlation to other cluster observables, especially those which trace mass (e.g., \citealt{benson/etal:2004,bonamente/etal:2008, mcinnes/etal:2009, marrone/etal:2010, melin/etal:2010, huang/etal:2010, plagge/etal:2010, culverhouse/etal:prep}). The SZ/X-ray and SZ/optical lensing correlations provide the mass calibration necessary for understanding the cluster mass function through cosmic time. By measuring the redshift evolution of the cluster mass function, constraints can be put on the matter density field and dark energy. Thus proper mass calibration is essential for studying cosmology with clusters \citep[e.g.,][]{francis/bean/kosowski:2005}. Finally, targeted SZ observations are expanding our knowledge of the physics of the ICM by probing the cluster gas pressure at large radius \citep{mroczkowski/etal:2009,plagge/etal:2010,komatsu/etal:prep} as well as at high resolution \citep{komatsu/etal:2001,mason/etal:2010}.

The first blind detections of galaxy clusters through their Sunyaev-Zel'dovich effect were reported by \citet{staniszewski/etal:2009} for the South Pole Telescope  (SPT; \citealt{carlstrom/etal:prep}).  Three of the clusters from \citet{staniszewski/etal:2009} were previously unknown.  \cite{hincks/etal:prep} reported on the cluster detections for the Atacama Cosmology Telescope (ACT; \citealt{fowler/etal:2007, swetz/etal:prep}). While the clusters from \cite{hincks/etal:prep} were previously known, they were blindly detected in the survey data. SPT and ACT have been able to realize blind detections from cosmologically significant surveys because they combine three essential design features: arcminute resolution matched to the size of clusters,  degree-scale fields of view for efficient surveying, and the unprecedented sensitivity of 1000-element bolometric detector arrays. In \cite{vanderlinde/etal:prep}, the SPT collaboration reported on twenty-one cluster candidates, twenty of which were optically confirmed \citep{high/etal:prep}.  In a follow-up study, \cite{andersson/etal:prep} presented the X-ray properties of the SPT sample and derived SZ/X-Ray correlations. Furthermore, \cite{vanderlinde/etal:prep} obtained cosmological constraints from an SZ-selected sample of clusters. 

\begin{figure*}[ht]
\begin{center}
  \resizebox{\textwidth}{!}{
    \plotone{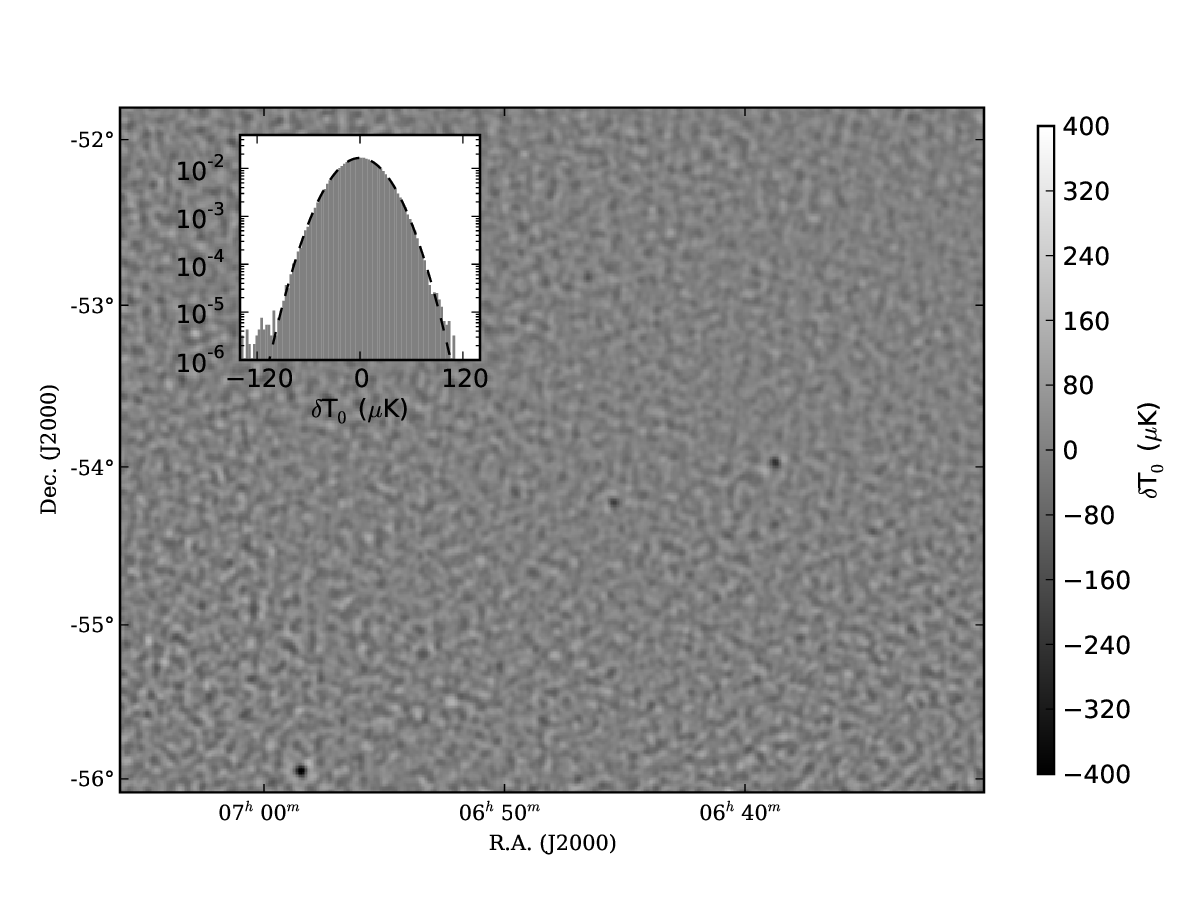}
  }
\caption{\arone\ sub-map.  The data have been weighted by a smooth function $\sqrt{N_{obs}/N_{obs,\,max}}$, where $N_{obs}$ is the number of data per pixel. This weighting levels the amplitude of white noise across the map to that corresponding to the deepest data. The data is then match-filtered with a $\beta$-profile ($\theta_{\rm c} = 1.0\arcmin$). The coverage becomes shallower toward the bottom-left of the map causing a visible increase in the rms. The inset shows the flux density distribution across the data weighted by the square-root of the number of data per pixel. The data distribution is shown as a gray histogram on which is plotted a dashed Gaussian distribution with standard deviation $24 \micro{\rm K}$. Sources were removed prior to filtering. Three clusters are recognizable from left to right: the Bullet Cluster (bottom-left), A3404 (middle), and AS0592 (middle-right). }
\label{fig:map}
\end{center}
\end{figure*}

In this work, the ACT collaboration reports on \numberOfClusters\ clusters detected in a subset of the data taken during the 2008 observing season. This paper describes the detection and millimeter properties of the clusters. Companion papers describe the physical properties and observed purity of the cluster sample \citep{menanteau/etal:prep} as well as cosmological implications \citep{sehgal/etal:prep}. These papers and others cited above are part of a larger initial science release from ACT which also includes results on the CMB power spectrum and associated parameter constraints \citep{fowler/etal:prep, das/etal:prep, dunkley/etal:prep, hajian/etal:prep} and on compact millimeter sources \citep{marriage/etal:prepa}.

This paper is organized as follows. The data reduction and catalog construction are described in Section \ref{sec:findingClusters}. The detections and their properties are discussed in Section \ref{sec:properties}. An estimate of purity and completeness of the sample is derived from simulations in Section \ref{sec:simulations}. Finally, in Section \ref{sec:scaling}, the SZ signal is compared to X-ray-derived mass for a high significance subset of the cluster sample.

\section{From Observations to Cluster Detection}
\label{sec:findingClusters}

The observations and methods to reduce the raw data to maps for this work are the same as used in the 2008 power spectrum study \citep{fowler/etal:prep} and extragalactic source study \citep{marriage/etal:prepa}. The primary difference between the source and cluster study lies in the compact signal profile used in the implementation of the matched filter-based detection. Therefore, we restrict our discussion here to a short summary of the observations and data reduction, and refer the reader to  \cite{fowler/etal:prep} and \cite{marriage/etal:prepa} for a more complete description. We describe in detail the particular implementation of the matched filter for clusters. For a description of the ACT receiver and instrumentation see \cite{swetz/etal:prep}.

\subsection{Observations and Data Reduction}

ACT is a six-meter telescope operating at 5200\,m in the Atacama Desert of northern Chile. The site was chosen for its excellent atmospheric transparency and access to both southern and northern skies. The telescope has three 1024-element arrays of transition edge sensors operating at \arone, \artwo, and \arthree. This study uses \arone\ data from a 455\,square-degree subregion of the 2008 southern survey. The subregion lies between right ascensions $00^{\mathrm h}12^{\mathrm m}$ and $07^{\mathrm h}08^{\mathrm m}$ and declinations $-56\degree11\arcmin$ and $-49\degree00\arcmin$. Figure \ref{fig:coverage} is a map of the sensitivity across the subregion along with the locations of the clusters reported in this study. The median rms of the map optimally filtered for detecting a $\beta$-model profile ($\beta=0.86$, $\theta_{\rm c}=1.00\arcmin$) is \medianNoiseOneArcminute.

\begin{figure*}[ht]
\begin{center}
  \resizebox{\textwidth}{!}{
    \plotone{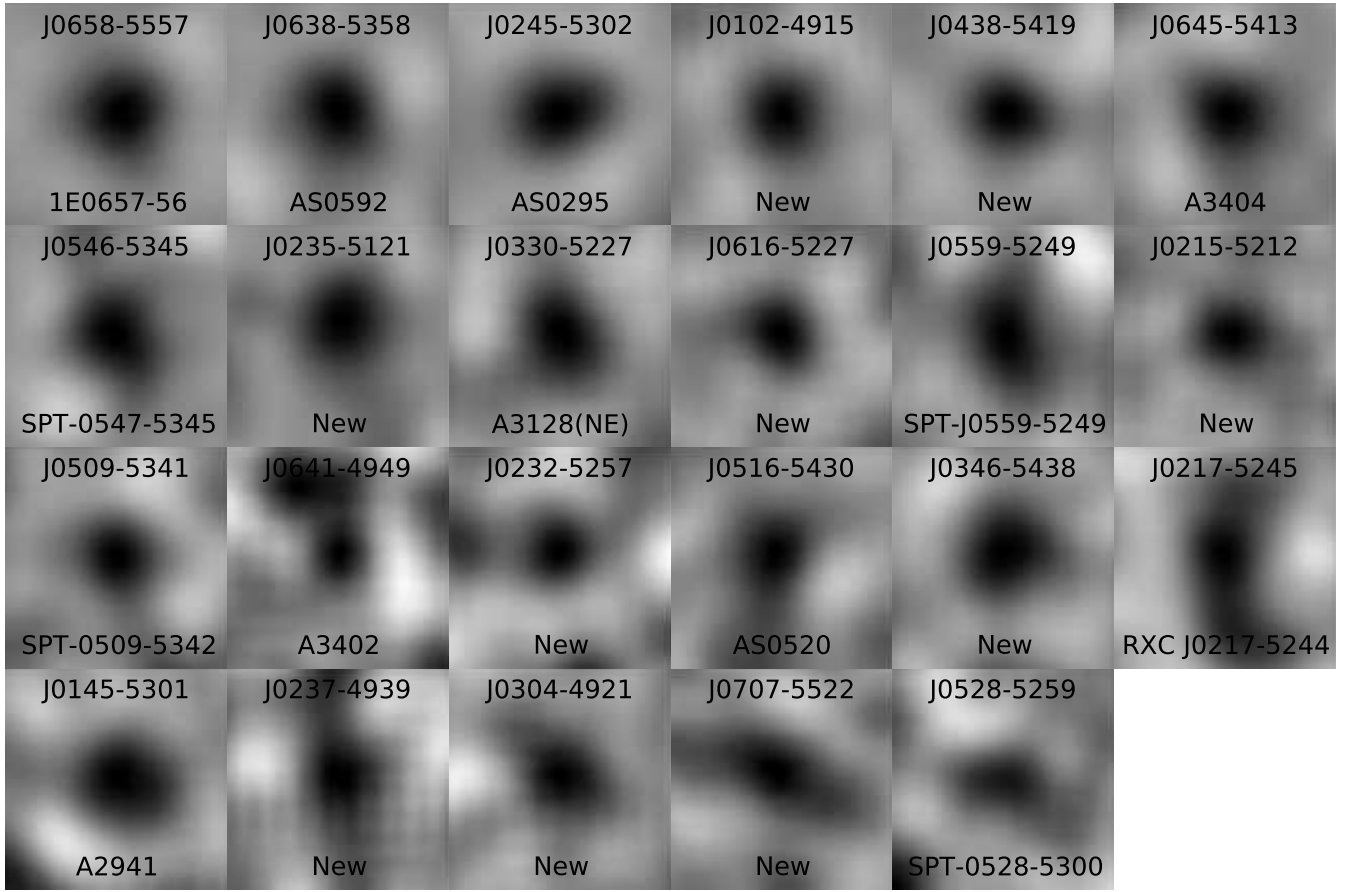}
  }
\caption{Temperature decrements at 148 GHz of \numberOfClusters\ clusters. Each thumbnail is approximately $8\arcmin$ on a side and is excised from the source-removed, filtered map. The grayscale of each thumbnail is symmetric about zero with a range set to the depth of the decrement. The ordering of the clusters is according to decreasing SNR from left to right and top to bottom. The structure northeast of ACT-CL J0641-4949 may be part of the cluster or may be contamination: the decrement of this nearby cluster may be noise boosted. ACT designations are at the top of each thumbnail and alternate names at the bottom, denoted ``New" if newly discovered by ACT.}
\label{fig:thumbnails}
\end{center}
\end{figure*}

The data for this study were calibrated to the temperature of Uranus with a precision of 6\%. The absolute positional uncertainty in the maps is established at $3.5\arcsec$ by comparison of ACT radio source locations \citep{marriage/etal:prepa} to cross-identified sources in the Australia Telescope 20\,GHz Survey \citep{murphy/etal:2010}. Note, however, that reported cluster locations have an additional uncertainty due to the effect of noise on the estimate of the cluster center. Considering the positional uncertainty for sources detected between signal-to-noise ratio (SNR) 5.5 and 10 reported in \citet{marriage/etal:prepa}, we expect estimates of the ACT cluster center positions to scatter with an rms of roughly 10$\arcsec$. Additional ambiguity in the cluster position arises in the case of an extended, non-spherical (e.g., disturbed) system. In such cases, the positional uncertainty can rise to arcminute scales.

With calibration and astrometry solved, the final step in the data reduction is map making. We iteratively solve for the maximum likelihood (ML) estimate of the map using a custom preconditioned conjugate gradient code \cite[e.g.,][]{press/teukolsky/vetterling:NRC:2e}. Because ACT samples the sky along multiple scan directions (i.e., the data is cross-linked), we are able to produce an unbiased ML map of the microwave sky with a faithful representation of structure from degree to arcminute scales.

\begin{figure*}[ht]
\begin{center}
 \resizebox{\textwidth}{!}{ \plotone{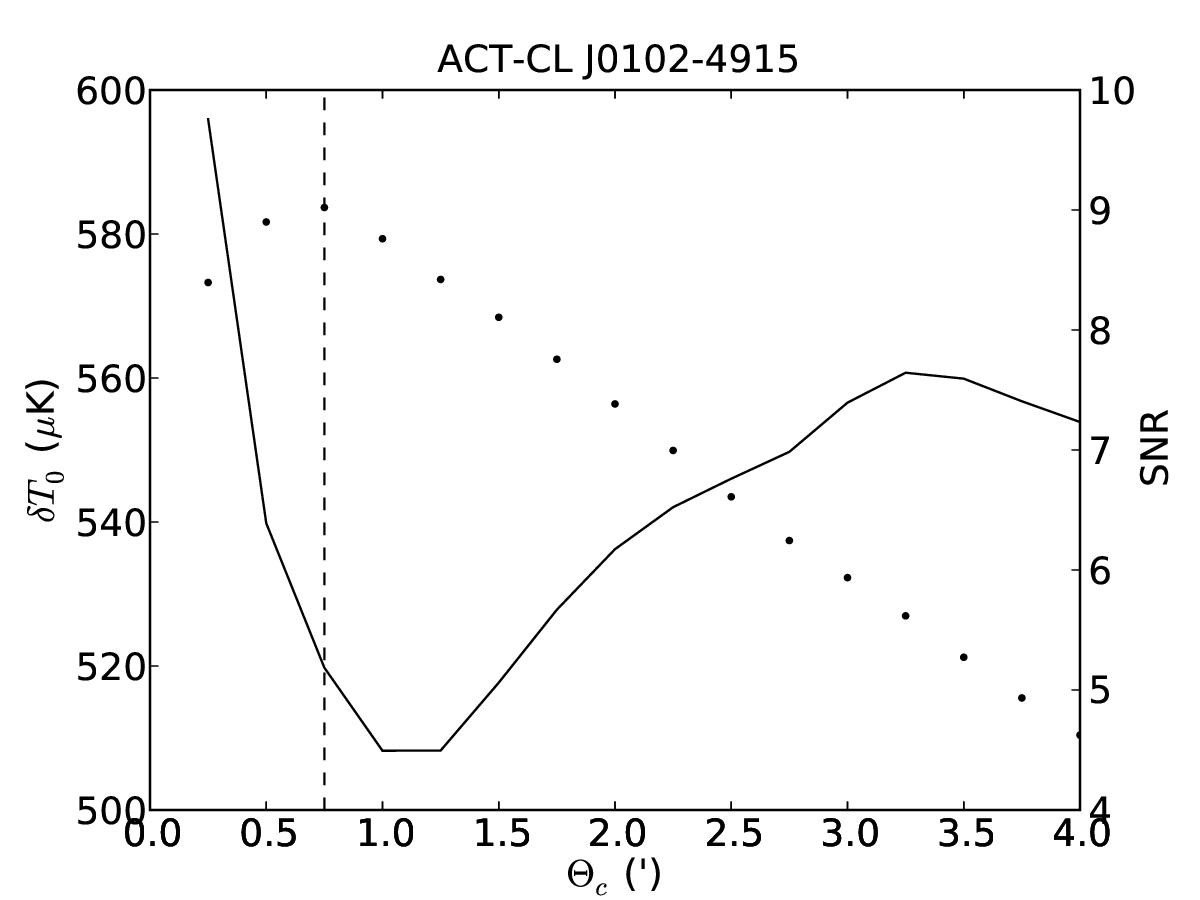}
\plotone{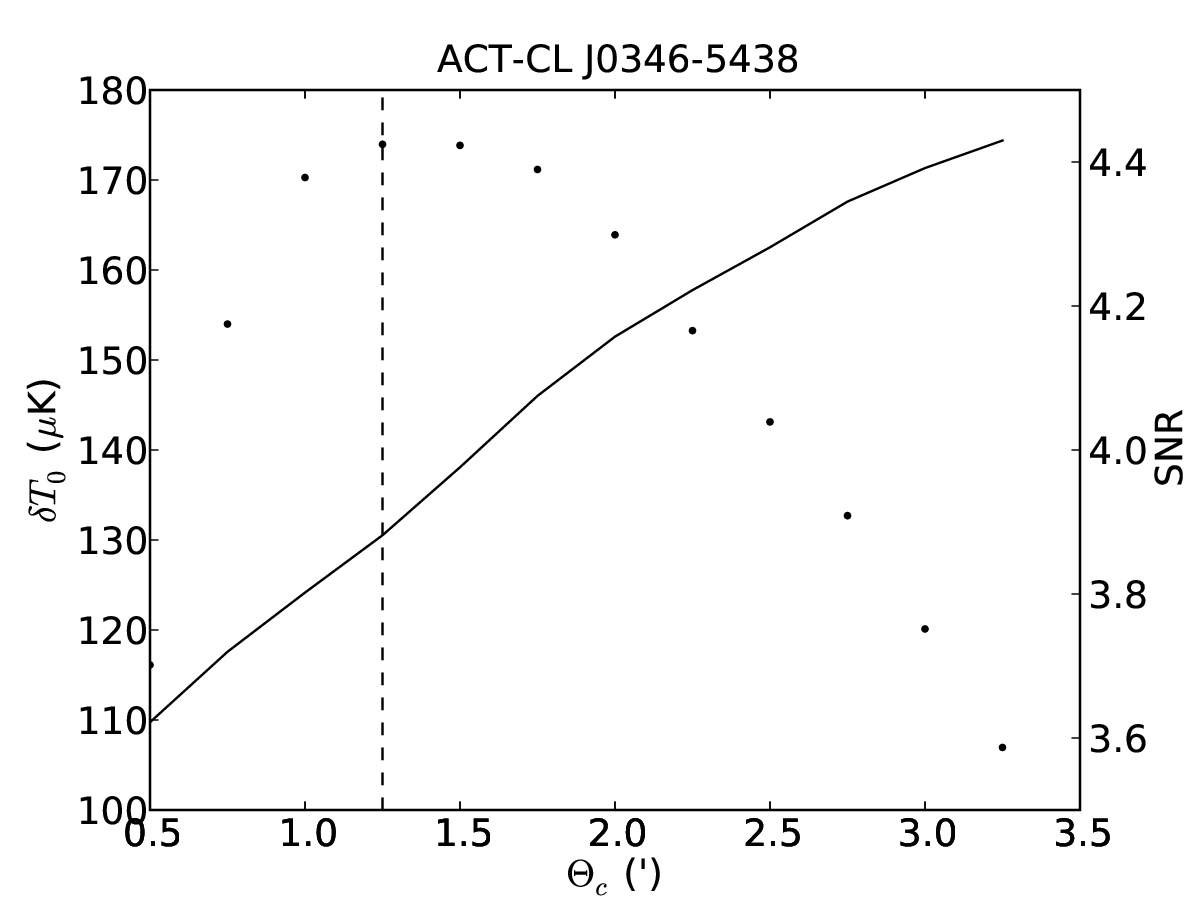}}
\caption{Decrement amplitude and SNR versus the filter core radius for two newly discovered clusters. The solid curve corresponds to the decrement while the dotted curve corresponds to the SNR. A vertical dashed line marks the maximum SNR identified by the extraction algorithm.}
\label{fig:detectionFunctions}
\end{center}
\end{figure*}

\subsection{Filtering and Cluster Extraction}
\label{sec:detection}

For cluster detection we use matched filters \citep{haehnelt/tegmark:1996,herranz/etal:2002,melin/bartlett/delabrouille:2006}. We model the sky temperature fluctuation at a point $\vec{x}$ as
\begin{equation}
\delta T(\vec{x}) = \sum_i \delta T_{0,i} B_{\theta_{\rm c},i}(\vec{x} - \vec{x_i}) + \delta T_{\rm other}(\vec{x})
\label{equ:model}
\end{equation}
where $\delta T_{0,i}$ and $B_{\theta_{\rm c},i}$ are the amplitude and unit-normalized profile of the $i^{\rm th}$ cluster. In what follows we choose $B_{\theta_{\rm c}}$ to be the  isothermal $\beta$-model ( $\beta=0.86$ ) with core radius $\theta_{\rm c}$ ranging from $0.25\arcmin$ to $4.0\arcmin$ and convolved with an isotropic ACT 148\,GHz beam from \cite{hincks/etal:prep}. The profile is tapered to zero by multiplication with a cosine in the range $0.5$--$5.5\theta_{\rm c}$. The choice of $\beta$ is motivated by the best fit to an average SZ profile in \citet{plagge/etal:2010} who found consistency between the $\beta$-model fit and the Generalized NFW profile fit to ${\rm Y}_{\rm X}$ in \citet{arnaud/etal:prep}. The temperature field $\delta T_{\rm other}(\vec{x})$ consists of noise modeled from difference maps, primordial CMB fluctuations with power spectrum from \citet{nolta/etal:2009}, as well as sources and undetectable (i.e., low mass) clusters. The spectral signature of the source and cluster contribution to $\delta T_{\rm other}(\vec{x})$ is modeled from fits to ACT data in \cite{fowler/etal:prep}. See \cite{marriage/etal:prepa} for a full discussion of $\delta T_{\rm other}(\vec{x})$. 

Before filtering, bright (${\rm SNR} > 5 $) sources are in-painted with sky temperature in the neighborhood of the source. Furthermore, the map is weighted by the inverse square root of the number of observations. This has the effect of flattening the white noise across the map. The map is filtered in the Fourier domain using  a matched filter:
\begin{equation}
\delta T_{\rm filt}(\vec{k}) = \frac{  \widetilde{B}_{\theta_{\rm c}}^*(\vec{k}) \mid \widetilde{\delta T}_{\rm other}(\vec{k}) \mid^{-2} \delta T(\vec{k}) } { \int   \widetilde{B}_{\theta_{\rm c}}^*(\vec{k'}) \mid \widetilde{\delta T}_{\rm other}(\vec{k'}) \mid^{-2} \widetilde{B}_{\theta_{\rm c}}(\vec{k'}) {\rm d}\vec{k'}}
\label{equ:filter}
\end{equation}
where $\widetilde{B}_{\theta_{\rm c}}(\vec{k})$ and $\widetilde{ \delta T}_{\rm other}(\vec{k})$ are the Fourier transforms of $B_{\theta_{\rm c}}$ and $\delta T_{\rm other}$, respectively. The map is filtered using $\beta$-models with core radii from $0.25\arcmin$ to $4.0\arcmin$ in $0.25\arcmin$ steps. These core radii were chosen to span the range of angular scales characterizing massive clusters from low to high redshift.  In a given map, the SNR of a detection is defined as the ratio of the extremum of the cluster decrement to the rms of the filtered map. The reported SNR for a given detection is the maximum SNR from the set of filtered maps. Figure \ref{fig:map} shows a subsection of the source-masked and filtered map containing three known clusters. Shown in an inset of this figure is the pixel flux distribution of the filtered map. The flux distribution suggests the filtered map is dominated by Gaussian noise.

The filter in Eq.~\ref{equ:filter} is normalized such that the central decrement in the filtered map is an estimate of the central decrement value in $\micro{\rm K}$. It is this value, recovered from the filtered map, which we use as our decrement estimate. As discussed in \cite{marriage/etal:prepa}, this estimate suffers from increased error and bias due to the 0.5$\arcmin$ pixel size of the ACT maps. In order to mitigate these errors, the pixel size of sub-maps around each detection is decreased by Fourier transforming the map and adding extra zeros to the boundary of the Fourier transform. The inverse of this zero-padded transform is then the original submap with smaller pixels. For this study, the pixel size is changed to $0.0625\arcmin$ which reduces the errors associated with the discrete approximation of Eq.~\ref{equ:filter} to less than 1\%. The errors we quote for the map-derived decrements are simply the decrement amplitude divided by the SNR.

\section{ Cluster Detections and Properties }
\label{sec:properties}

Table \ref{tab:clusters} lists clusters detected by ACT. All detections in the table have been  confirmed through optical follow-up imaging of a larger catalog of all SZ candidates from the ACT data \citep{menanteau/etal:prep}. The table is divided into two groups of candidates: candidates detected at SNR~$>$~\highSNR\ and candidates detected at lesser significance. The optical  follow-up showed that for candidates detected at SNR~$>$~\highSNR, the larger catalog of all SZ candidates is 100\% pure (See also Section \ref{sec:simulations}.).  Of the \numberOfHighSNRClusters\ detections at SNR~$>$~\highSNR, \numberOfNewHighSNRClusters\ are newly discovered systems. Among the \numberOfLowSNRClusters\ candidates confirmed at lower SNR, \numberOfNewLowSNRClusters\ are new detections. Thumbnails of the cluster decrements are shown in Figure \ref{fig:thumbnails}.

The multifrequency properties of the clusters covered in this work are discussed at length in \citet{hincks/etal:prep} and \citet{menanteau/etal:prep}. Paraphrasing these studies, X-ray luminosities were computed from Chandra and XMM observations where available and supplemented by ROSAT
All Sky Survey for the remaining clusters.  These luminosities suggest that the full sample is composed of high mass clusters with a characteristic mass of $M_{500c} \sim 6 \times 10^{14} M_\odot$, where $M_{500c}$ is defined as the mass of the cluster at five hundred times the critical density of the universe at the cluster
redshift. The sample has a median redshift of 0.44. All but one of the clusters at $z > 0.35$ in the sample were discovered via the SZ.

Of the new ACT clusters, the most notable is ACT-CL J0102-4915. This cluster has a decrement amplitude ($\delta T'  = -1046\pm116 \mu K$) comparable to that of the Bullet Cluster ($\delta T' = -734\pm64 \mu K$) as well as a smaller angular core radius, consistent with its higher redshift ($z_{\rm photo} = 0.75$). The X-ray luminosity (0.1--2.4 keV) derived from ROSAT observations is $18.39 \pm 6.49 \times 10^{44}$~erg~s$^{-1}$ which is also comparable to that of the Bullet as well as the other massive clusters in our sample \citep{menanteau/etal:prep}. The cluster richness is also consistent with that of a massive cluster ($N_{gal}\approx50$).  The strength of this decrement may result from merger activity, elongation along the  line of sight, or other decrement-enhancing effects. Planned spectroscopic and IR follow-up observations will determine the true nature of the decrement.

Table \ref{tab:clusters} contains estimates of the cluster decrement amplitudes $\delta T_0$  based upon the matched filter described in Section \ref{sec:detection}. Also listed is the core radius $\theta_c$ corresponding to the filter that exhibits the largest SNR for each cluster. Figure \ref{fig:detectionFunctions} shows the temperature decrement and SNR of two newly discovered clusters plotted as a function of filter scale (i.e., the core radius). For some of the clusters, such as ACT-CL J0346-5438 shown in the figure, the SNR is a relatively broad function of core radius. The corresponding uncertainty in the size of the cluster results from the blending of the cluster decrement with the background (e.g., CMB). This confusion for single-frequency data was observed by \citet{melin/bartlett/delabrouille:2006} as well as \citet{vanderlinde/etal:prep}. For clusters with a more compact and pronounced core, like ACT-CL J0102-4915, the cluster-background confusion is smaller and so the optimal filter scale is easier to determine from the data. Confusion with the background inflates $\theta_c$ while contamination by detector white noise causes a cluster to appear more compact than it actually is. The latter effect can be explained by the fact that most of the power in the white noise field is at the smallest scales. Simulations (See Section \ref{sec:simulations}.) show contamination for clusters detected outside the range $\theta_c = 0.75\arcmin-1.75\arcmin$. For this reason, $\theta_c$ estimates reported below $0.75\arcmin$ or above $1.75\arcmin$ are suspect. Incorporation of ACT's multiple frequencies will help distinguish clusters from background. Additional years of data will help reduce the fraction of white-noise-boosted detections.

The $\delta T_0$ recovered from the filter is the amplitude of the cluster decrement in the ACT map. As such, it is the amplitude of the intrinsic cluster profile convolved with the ACT beam. In Table \ref{tab:clusters} we also include an estimate of the decrement $\delta T_0'=\delta T_0C^{-1}(\theta_{\rm c})$ where we have corrected the raw $\delta T_0$ by the smearing effect of the beam on the height of the corresponding $\beta$-profile. This correction for a $\beta$-profile with $\theta_{\rm c}=1\arcmin$ is 1.67. The correction is larger for smaller core radii. To estimate  the uncertainty in the recovered $\delta T_0'$, the cluster extraction algorithm was run on a simulation  with realistic ACT noise, CMB and clusters with symmetric $\beta$-model profiles. The more realistic simulations of \cite{sehgal/etal:2010} were not  used as they were not provided an estimate of the decrement amplitude. As the $\beta$-model profiles in this simulation are matched to our filters, the error estimate from these simulations should be taken as a lower limit. With those caveats, the mean error on $\delta T_0'$ from the simulations is 20\% greater than the statistical error inferred from the SNR (i.e., systematic errors are 40\% of the total error budget). This modest increase in the error budget was accompanied by a large scatter in recovered scale ($\sim$30\% with respect to the input scale value) with a non-Gaussian tail towards larger recovered radii. This implies that while the physical scale of the cluster may be confused with the background the amplitude of the peak is relatively immune to this confusion. When the cluster extraction algorithm was run on the $\beta$-model simulations with prior knowledge of the scale, the resulting scatter in recovered $\delta T_0'$ values was consistent with the statistical error inferred from the SNR. Thus we infer that, in the ideal case of $\beta$-model cluster profiles, uncertainty in scale adds $20\%$ scatter to our uncertainty in $\delta T_0'$. For more realistic cluster profiles (e.g., asymmetric), this scatter should be larger.

Because the cluster counts are a falling function of mass (and there for of decrement amplitude), clusters detected at low significance are more likely to be lower mass systems with a noise-boosted decrement than a higher mass system with a noise-suppressed decrement. In particular, members of the lower significance (SNR $<$ 6) fraction of the ACT detections are likely to have their $\delta T_0'$ overestimated due to noise-boosting. In this work, no correction has been made for this bias. An additional bias arises from optimizing the detection significance over the two celestial coordinates and $\theta_c$. \cite{vanderlinde/etal:prep} show that this latter bias, when referred to SNR, can be shown to be $\sqrt{SNR^2 - 3}$. In the case of SNR~=~4, therefore, the optimization over location and core radius increases the raw decrement estimate  by 10\% on average.

Decrement measurements at 150 GHz have been published for a number of the known clusters in the current ACT sample. \citet{halverson/etal:2008} using the APEX telescope reported  a decrement for the Bullet Cluster $-771\pm71 \micro {\rm K}$ while \citet{plagge/etal:2010} using the South Pole Telescope reported $-932 \pm 43 \micro {\rm K}$. \cite{plagge/etal:2010} also found decrements for  AS 0520 ($-217 \micro$K), A 3404 ($-472 \micro$K), AS 0592 ($-529 \micro$K), and RXC J0217.2-5244 ($-217 \micro$K) all of which are in reasonable agreement with the $\delta T'_0$ values in Table \ref{tab:clusters}. The central values of the SPT estimates are, however, systematically more negative than the central values reported in Table \ref{tab:clusters}. 

Table \ref{tab:clusters} also contains the central Compton $y$ parameter $y'_0$ which is related to the central decrement $\delta T'_0$ by :
\begin{equation}
  \frac{\delta T'_0}{T_{CMB}} = y'_0 \left(x\frac{e^x + 1}{e^x - 1} - 4\right)(1+\delta_{SZE}(x, T_e)).
  \label{equ:sz}
\end{equation} 
In this equation $x = h\nu/k_BT_{CMB}$ and $T_e$ is the temperature of the electron gas. In this equation, we have omitted the $\sim10\%$ contribution of the kinetic Sunyaev-Zel'dovich (kSZ) effect which arises due to the peculiar velocity of the cluster \citep{sunyaev/zeldovich:1970}. Because we do not know the bulk velocity of the clusters, we do not account for the contribution of the kinetic effect which adds to the uncertainty in the derived $y$ parameter. The $y$ parameter itself is the product of the optical depth through the cluster and $k_{\rm B}T_e/ m_e c^2$ where $m_e$ is the electron mass:
\begin{equation}
y'_0=\frac{ k_B T_e }{ m_e c^2} \int n_e \sigma_T d\ell  
\end{equation}
where $\sigma_T$ is the Thomson cross section and isothermality of the cluster is assumed.
The relativistic correction  $\delta_{SZE}(x, T_e)$ at \arone\ varies from -0.03 to -0.10 for the ACT sample, corresponding to cluster temperatures in the range 5--15 keV \citep{nozawa/etal:2006}. We have used X-ray temperatures  tabulated in \cite{menanteau/etal:prep}  as input to the relativistic correction for a subset of the clusters. For clusters without an X-ray temperature, we use a correction of -0.05 which roughly corresponds to a temperature of 7.5 keV, characteristic of our sample.

For all but one case, there is no obvious correlation of the cluster decrements with strong sources. The one exception is Abell S0295.  Approximately 4$\arcminute$ north-east of the cluster is ACT-S J024539-525756 \citep{marriage/etal:prepa} with a \arone\ flux density of 18.4~mJy (SNR=7.6). The source is also identified at 0.84\,GHz in the Sydney University Molonglo Sky Survey (SUMSS) catalog \citep{mauch/etal:2003}. Therefore the source flux density is most likely dominated by synchrotron emission. From simple geometric arguments, there is approximately a 10\% chance that one of the the ACT sources would randomly fall within 4$\arcmin$ of one of the twenty-three clusters, so the source-cluster proximity may be a projection effect. Furthermore, the source's large angular displacement from the cluster implies that it is unlikely a lensed background galaxy. Finally, note that this source does not appear near Abell S0295 in Figure \ref{fig:thumbnails} because the source was removed from the data prior to filtering.

There are slight differences between the cluster properties as presented in Table \ref{tab:clusters} and in \citet{menanteau/etal:prep}. The SNR and, to a lesser degree, position differ due primarily to two facts. First, the data reduction (i.e., the map) used for  \cite{menanteau/etal:prep} was a preliminary version of that used for this work. The map used for this work corresponds to that used in \cite{fowler/etal:prep} and \cite{marriage/etal:prepa}. Second, the signal and noise terms used in the matched filter for the candidate selection in  \cite{menanteau/etal:prep} are different from those used in this study. These differences in analysis arose from the need to execute the optical follow-up before the ACT pipeline and cluster extraction algorithm were fully mature. The differences in SNR generally correspond to approximately $1 \sigma$ fluctuations between the reductions.

\section{Simulated Purity and Completeness}
\label{sec:simulations}

In order to assess the ratio of true detections to total detections (purity) and the fraction of clusters recovered (completeness), we run our filtering and extraction algorithm on simulations of the ACT data.   The simulations were constructed using the millimeter signal (SZ, CMB, etc.) from \citet{sehgal/etal:2010} and noise from ACT difference maps. In order to avoid confusion due to source-SZ correlations included in \citet{sehgal/etal:2010}, the source component of those simulations was not included. The undetected/unmasked sources in the data contribute a small fraction of the rms of the filtered map which is dominated by white detector and photon shot noise \citep{marriage/etal:prepa}. 

\begin{figure}[ht]
\begin{center}
\resizebox{\columnwidth}{!}{\plotone{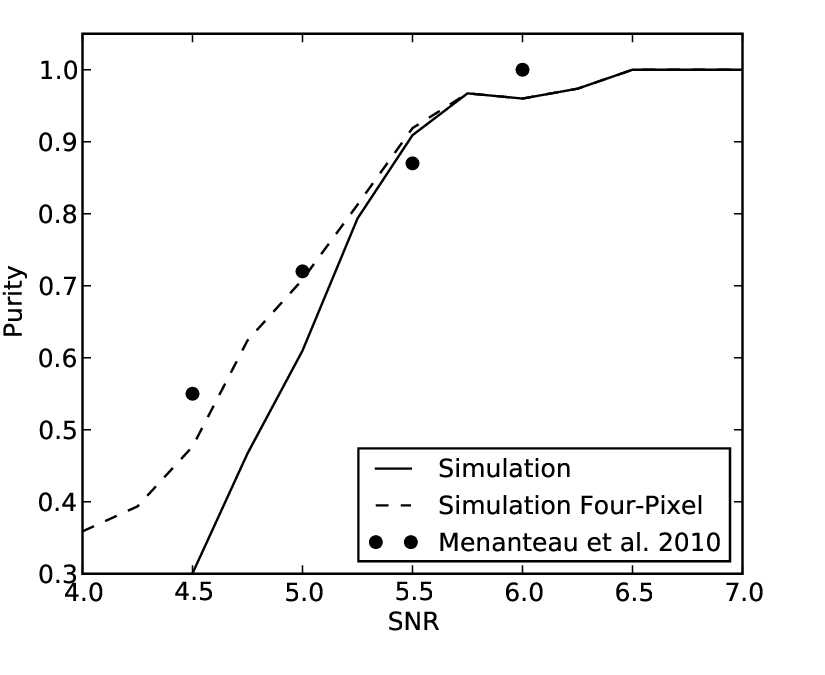}}
\caption{Simulated purity. Purity is defined as fraction of true detections in the sample. The four-pixel candidate selection, which was used to produce the sample in \cite{menanteau/etal:prep}, requires cluster candidates to have at least four pixels with SNR~$>$~4.0. The dashed line gives the simulated purity associated with the four-pixel selection function. The solid line is the purity function corresponding to a selection function based on the single highest SNR cluster pixel. As expected the four-pixel purity of the simulated sample is in reasonable agreement with that derived from optical follow-up of candidates from the data \cite{menanteau/etal:prep}. The sample is essentially  pure at SNR~$>$~\highSNR.}
\label{fig:purity}
\end{center}
\end{figure}
\begin{figure}[ht]
\begin{center}
\resizebox{\columnwidth}{!}{\plotone{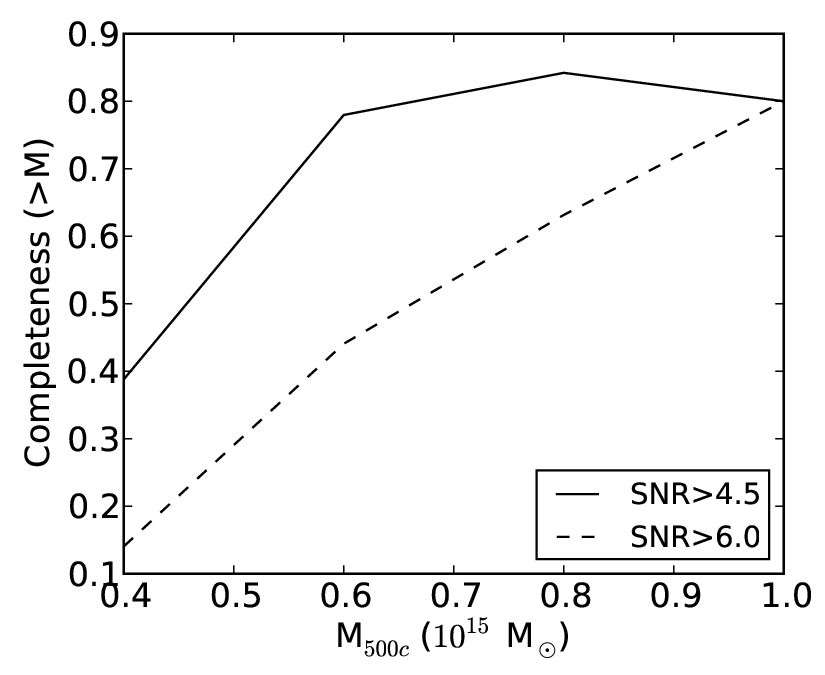}}
\caption{Cumulative completeness of the simulated sample. At \completenessRange the approximation to the total sample (SNR~$>$~4.5) is \totalSampleCompleteness\ complete. The completeness degrades at the edges of the map: taking the cleanest 230 sq-deg from \cite{fowler/etal:prep}, the completeness rises to 90\%.}
\label{fig:completeness}
\end{center}
\end{figure}

Although it is expected that, from a noise power perspective, omitting the sources in this study will not significantly impact the analysis, there remains  concern that millimeter sources may be correlated with clusters, either as cluster members or as lensed backgrounds. Studying 573 nearby clusters with 1.4 GHz data from the NRAO VLA Sky Survey, \cite{lin/mohr:2007} found that the fraction of BCGs being radio-loud is higher compared to that of cluster galaxies of similar luminosity. Extrapolating in frequency and redshift, this study concluded that roughly 10\% of clusters may have 150 GHz decrements ``filled in'' by  synchrotron emission from resident AGN. \citet{marriage/etal:prepa} included a cross-comparison of the 150 GHz bright source population detected by ACT and the 23 REFLEX X-ray clusters \citep{bohringer/etal:2004} which fall within the survey area. The study cross-identified three clusters with 150 GHz sources. These three clusters were all at redshifts below $z=0.06$ in contrast to the higher redshift sample of this work.  Lensed background dusty star-forming galaxies may also contaminate the SZ signal. \cite{wilson/etal:2008} found a bright, lensed, dusty source at 1.1 mm behind the Bullet Cluster. \cite{vieira/etal:2010} found that the lensed dusty population were characterized my maximum 150 GHz flux densities below 5 mJy range. This finding is consistent  with the \cite{wilson/etal:2008} bright Bullet Cluster lensed source flux of 15.9~mJy at 1.1~mm ($\sim 4$~mJy at 150 GHz assuming a spectral index of 2). Filtered at characteristic scale (.75') of this work, such sources would produce a peak brightness of 10 $\mu$K.  As the decrements in this work measure of order 100 $\mu$K, contributions from dusty sources may affect the purity and completeness of faint end of the sample, but the bright end (SNR $>$ 6) should be relatively immune to detection problems due to dusty sources.

In order to build simulated catalogs to compare to that from the data, we run the filtering and detection pipeline described in Section \ref{sec:detection} on each of six simulations extracted from the \citet{sehgal/etal:2010} sky. 
We identify decrements with clusters from the simulated catalog with $M_{500c}~>$~\minimumMass\ and at redshifts \redshiftRange. The identification radius is \associationRadius. This radius was chosen by studying significant decrements detected in the simulations and their offset from the position reported in the simulated catalog. By coincidence, this is also approximately the size of the optical fields of view used in the \cite{menanteau/etal:prep} follow-up work.  In nearly all cases, the arcminute scale displacements resulted from extended non-spherical, often double peaked, clusters. In these cases it was easier for noise fluctuations to move the detection position away from the minimum of the cluster potential as reported in the catalogs from \cite{sehgal/etal:2010}.

Figure \ref{fig:purity} shows the purity as derived from the simulations. In order to reproduce the original selection function from \cite{menanteau/etal:prep} we require that a cluster candidate have four conjoined map pixels with SNR~$>$~4.0. This requirement is intended to boost the purity of the sample with respect to a sample drawn with the requirement that only a single pixel have a high SNR. Figure \ref{fig:purity} shows the purity versus SNR for both of these selection effects. As expected, the four-pixel requirement increases the purity as determined by the simulations. Furthermore the four-pixel purity agrees reasonably well with the purity measured in \cite{menanteau/etal:prep}.  The simulations imply that the purity of the sample with SNR $>$ 6 is \highSNRPurity, consistent with the finding that all these clusters have optical counterparts. Furthermore, both the four-pixel and single-pixel selection process yield the same purity for the high SNR sample. Below SNR~=~\highSNR, simulations show that the four-pixel purity falls off in a manner consistent with the purity determined from the optical follow-up. The single-pixel purity falls significantly faster with decreasing SNR compared to that of the four-pixel or optically-measured purity. We therefore choose the four-pixel selection function to evaluate the completeness of the sample below.

It is straightforward to simulate the cumulative completeness as a function of cluster mass for the pure population at SNR~$>$~\highSNR. As shown in Figure  \ref{fig:completeness} with the dashed curve, we count the simulated clusters above a given mass and detected at SNR~$>$~\highSNR. We then divide the number of detected clusters by the total number of clusters in the same mass range. The resulting cumulative completeness shows that the high SNR sample is approximately \highSNRCompleteness\ complete for \completenessRange\ and rises to 80\% complete for  $M_{500c} > 1.0\times10^{15}M_\odot$. One reason for incompleteness among the high mass population is that clusters at $z<0.3$ are characterized by larger angular scales than their high-redshift counterparts and start blending with the fluctuations in the cosmic microwave background. Dividing the simulated sample in redshift bins of $\delta$z~=~0.2 bins between z~=~0.1 and z~=~1.1, the completeness of this sample, in the mass range \completenessRange, increased smoothly from 30\% in the lowest redshift (z~=~0.1--0.3) bin to 70\% complete in the highest bin (z~=~0.9--1.1).

It is not so straightforward to reproduce the cumulative completeness function for the entire sample presented in \cite{menanteau/etal:prep} which is based on a previous reduction of the data. As shown in Figure 13 of  \cite{menanteau/etal:prep}, the follow-up observations were completed for a preliminary candidate list down to SNR~=~4.2. Below SNR~=~4.8, the follow-up was performed for a decreasing fraction of candidates such that, by SNR~=~4.2, the fraction of the total candidates followed up was 50\%. In order to approximately simulate the smoothly declining selection function of the follow-up for candidates in the range SNR~=~4.2-4.8,  we simply introduce a hard cut at SNR~=~4.5 where the sampling goes from 100\% to 0\%.  

With this prescription for modeling the entire sample, the simulations suggest that \totalSampleCompleteness\ of all clusters with  \completenessRange\ are detected. Figure \ref{fig:completeness} shows the full cumulative completeness function with a solid curve.  Unlike in the case of the high SNR sample discussed above, the simulations suggest that completeness of the full sample in the range \completenessRange\  is not a strong function of redshift (z~=~0.1--1.1).  Another important reason for the incompleteness is that a fraction of massive clusters are disturbed, showing elongated structure in the plane of the sky, and thus are not well matched to our $\beta$-model filter. Conversely, the simulations suggest our sample likely includes high-redshift relaxed, spherical clusters of lower mass (e.g., $4\times10^{14} M_\odot$). Another feature of the data which drives the overall completeness down is the high level of noise towards the edge of the map. If the region under consideration is restricted to the central 230 square degrees used for the power spectrum study in \cite{fowler/etal:prep}, then the completeness estimate for clusters with \completenessRange\ increases to $\sim$90\%.

\section{SZ-Mass Correlation}
\label{sec:scaling}

In this section, we perform a largely illustrative, preliminary study of the correlation of the SZ signal with mass for the more significant (SNR~$>$~5) half of the SZ detections. For the SZ signal we use the central $y_0'$ as described in Section \ref{sec:properties} and listed in Table \ref{tab:clusters}. For the total mass, we derive $M_{500c}$ using X-ray luminosities from \cite{menanteau/etal:prep} and the $L_x - M$ relation from \cite{vikhlinin/etal:2009}. For all clusters but Abell 3128(NE), the X-ray luminosities were derived from the ROSAT All Sky Survey. For Abell 3128(NE), the available higher resolution Chandra results from \cite{menanteau/etal:prep} were used to help separate the flux from the more massive background cluster from the less massive cluster at low redshift \citep{werner/etal:2007}.  An additional scale factor of $\sim1.6$ was needed to convert the luminosities from the 0.1-2.4 keV band in \cite{menanteau/etal:prep} to the 0.5-2.0 keV band required by the correlation in \cite{vikhlinin/etal:2009}.  The resulting $M_{500c}$ estimates are shown in Table \ref{tab:act_cluster_masses}. The error given for $M_{500c}$ only includes statistical error propagated from error on $L_X$. Additional scatter about this relation was measured to be 48\% \citep{vikhlinin/etal:2009}. due to evolution and non-trivial processes in the ICM as well as  Ongoing optical and X-ray follow-up of the ACT cluster sample will improve these mass estimates in the near future.

\begin{figure}[ht]
\resizebox{\columnwidth}{!}{\plotone{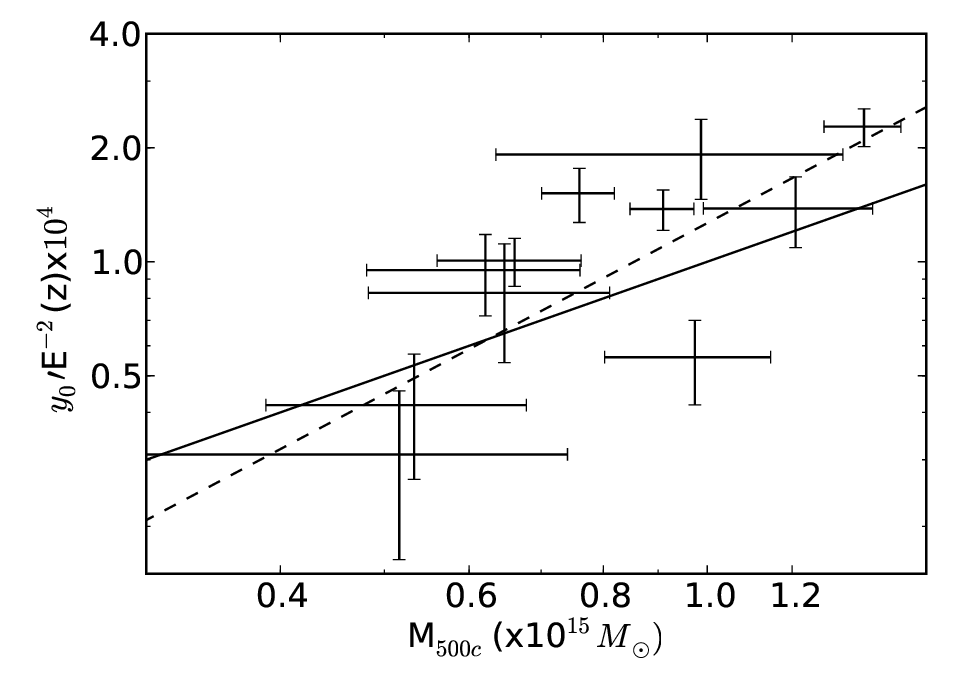}}
\caption{SZ-Mass Correlation. Estimated Compton $y$ parameter estimates (Table \ref{tab:clusters}) are plotted, scaled for self-similar evolution, against X-ray luminosity-derived mass estimates  (Table \ref{tab:act_cluster_masses}) for the more significant half (SNR~$>$~5) of the cluster sample. The line plotted is the self-similar model $y_0' \propto M E^2(z)$ (See Appendix \ref{app:scaling}.). The model, which is not fit to the data, agrees reasonably well with the ACT data. For comparison, the dashed line shows a simulation-motivated model with logarithmic slope 3/2 \citep{motl/etal:2005}. }
\label{fig:YvsM}
\end{figure}

For self-similar evolution, the central  $y_0'$ is expected to scale as $ME^2(z)$ where $E^2(z)~=~\Omega_M(1~+~z)^3~+~\Omega_\Lambda$ for a flat $\Lambda$CDM universe.  The conventional argument behind this scaling is given in Appendix \ref{app:scaling}. Anticipating such a scaling, in Figure \ref{fig:YvsM} we plot  $y_0'$ values scaled by $E^{-2}(z)$ against their corresponding $M_{500c}$ estimates. Values used for $\Omega_M$ and $\Omega_\Lambda$ are 0.26 and 0.74, respectively \citep{komatsu/etal:prep}. The errors plotted for $y_0'$ are purely statistical. As discussed in Section \ref{sec:properties}, systematic errors, primarily due to uncertainty in the core radius, would increase the errors on $y_0'$ by at least 20\%. Because of systematic uncertainties associated with the estimators for $y_0'$ and $M_{500c}$ no attempt has been made to fit scaling relations to these data. Given this precaution, however, the data appear consistent with the self-similar model plotted as a solid line in the figure. An additional dashed line with $y_0' \propto M^{3/2} E^2(z)$ approximates the $y_0-M$ relation from numerical simulations including star formation and supernova feedback from \cite{motl/etal:2005}. The data also appear consistent with this model.

The SZ-Mass relation presented here is for illustrative purposes using central $y$ parameter estimates and a mass proxy that are available now for all clusters but which have a large scatter. The SZ signal estimate will be improved in future analyses using additional ACT bands as well as additional information on cluster gas and temperature profiles from the X-ray \cite[e.g.,][]{andersson/etal:prep}. The precise relationship between the SZ signal and mass needs to be determined from more precise mass proxies such as X-ray temperature, galaxy velocity dispersion, and weak lensing.  This work is in progress and will be reported in future publications.

\section{Conclusion}

ACT has identified the SZ signatures of \numberOfClusters\ optically-confirmed galaxy clusters at a frequency of \arone\ in data taken during the 2008 observing season. Of these clusters, ten are new discoveries. Together, ACT and SPT have reported twenty-five newly discovered systems. The Planck Satellite will soon add its complementary detections to the growing sample of SZ cluster discoveries. Due to the unique nature of the SZ effect, the sample is biased towards high-mass clusters at high redshifts. As such the sample complements current X-ray and optically-selected catalogues. Among the clusters discovered by ACT,  ACT-CL J0102-4915 stands out as a high redshift cluster having an SZ temperature decrement comparable to that of the most massive known clusters. Purity estimates from simulations of the cluster extraction agree with the sample purity derived from optical follow-up in \cite{menanteau/etal:prep}. The simulations also estimate completeness of the sample at roughly 80\% for $M_{500c} > 6 \times 10^{14} M_\odot$. Over the same mass range, the completeness increases to greater than 90\% if only the lowest noise half of the survey area is considered. A comparison of the clusters' central Compton $y$ parameters with X-ray luminosity-derived masses is in agreement with both self-similar scaling as well as scalings based on simulations which include star formation and supernova feedback. 

Future work with the SZ effect from galaxy clusters will include ACT data from multiple seasons and multiple millimeter bands. In addition to the southern survey described in this work, ACT is also surveying the celestial equator to complement extant and future multifrequency cluster work in this highly accessible strip of sky.

\acknowledgments

The ACT project was proposed in 2000 and funded by the U.S. National Science Foundation on January 1, 2004. Many
have contributed to the project since its inception. We especially
wish to thank Asad Aboobaker, Christine Allen, Dominic Benford, Paul
Bode, Kristen Burgess, Angelica de Oliveira-Costa, Sean Frazier, Nick Hand, Peter Hargrave,
Norm Jarosik, Amber Miller, Carl Reintsema, Felipe Rojas, Uros Seljak, Martin
Spergel, Johannes Staghun, Carl Stahle, Max Tegmark, Masao Uehara,
Katerina Visnjic, and Ed Wishnow. It is a pleasure to acknowledge Bob
Margolis, ACT's project manager. Reed Plimpton and David Jacobson
worked at the telescope during the 2008 season. Naoki Itoh and Satoshi Nozawa provided 
code for calculating relativistic corrections to the SZ. ACT operates in the Parque Astron\'omico Atacama in
northern Chile under the auspices of Programa de Astronom'a,
a program of the 
Comisi\'on Nacional de Investigaci\'on Cient\'ifica y Tecnol\'ogica  (CONICYT).

This work was supported by the U.S. National Science Foundation
through awards AST-0408698 for the ACT project, and PHY-0355328,
AST-0707731 and PIRE-0507768. Funding was also provided by Princeton
University and the University of Pennsylvania.  The PIRE program made
possible exchanges between Chile, South Africa, Spain and the US that
enabled this research program.  Computations were performed on the GPC
supercomputer at the SciNet HPC Consortium.  SciNet is funded by: the
Canada Foundation for Innovation under the auspices of Compute Canada;
the Government of Ontario; Ontario Research Fund -- Research
Excellence; and the University of Toronto.

TM was supported through NASA grant NNX08AH30G. JBJ was supported by the 
FONDECYT grant 3085031. ADH received additional support from a Natural 
Science and Engineering Research Council of Canada (NSERC) PGS-D scholarship. 
AK and BP were partially supported through NSF AST-0546035 and AST-0606975,
respectively, for work on ACT\@.  HQ and LI acknowledge partial support
from FONDAP Centro de Astrof\'isica. NS is supported by the U.S. Department of Energy contract 
to SLAC no. DE-AC3-76SF00515. RD was supported by CONICYT,
MECESUP, and Fundaci\'on Andes. RH was supported by the Rhodes Trust.  
ES acknowledges support by NSF Physics Frontier Center grant PHY-0114422 to the Kavli Institute of
Cosmological Physics. YTL acknowledges support from the World Premier
International Research Center Initiative, MEXT, Japan.  The ACT data will
be made public through LAMBDA (http://lambda.gsfc.nasa.gov/) and the
ACT website (http://www.physics.princeton.edu/act/).

%
%

\clearpage

\begin{landscape}

\begin{deluxetable}{cccrrccccc}

\tabletypesize{\footnotesize}
\tablecolumns{10} 
\tablecaption{ACT 2008 \arone\ SZ-Selected Galaxy Clusters}
\tablehead{
           \colhead{ACT ID}                                    &
           \multicolumn{2}{c}{RA  (J2000)   Dec}               &
           \colhead{SNR}               &
           \colhead{$\theta_{\rm c}$}    &               
           \colhead{$\delta T_0$\tablenotemark{a}}      &
            \colhead{$\delta T'_0$\tablenotemark{b}} &
            \colhead{$y'_0$} &
            \colhead{$z$\tablenotemark{c}} &
           \colhead{Alternative ID}                   
            \\
           \colhead{}                                          &        
           \colhead{\phn{h}\phn{m}\phn{s}}                         &
           \colhead{\phn{\arcdeg}~\phn{\arcmin}~\phn{\arcsec}} &
           \colhead{}                                          &        
           \colhead{($\arcmin$)}                                          &            
           \colhead{ ($\micro{\rm K}$)}             &
           \colhead{ ($\micro{\rm K}$)} & 
           \colhead{ ($\times10^{-4}$) } &   
           \colhead{}        &
           \colhead{}                             
           }
\startdata
\multicolumn{10}{c}{All Candidates SNR $>$ \highSNR} \\ \\
ACT-CL J0658-5557 &  06:58:30  &  $-55$:57:04  &  11.5  &  1.25  &  $ -483 \pm 42 $  & $ -734 \pm 64$ &    $ 2.98 \pm 0.26$ &   0.30  & 1E0657-56 (Bullet) \\ 
ACT-CL J0638-5358 &  06:38:46  &  $-53$:58:45  &  10.0  &  1.25  &  $ -278 \pm 28 $  & $ -423 \pm 42$ &    $ 1.68 \pm 0.17$ &   0.22  & AS0592 \\ 
ACT-CL J0245-5302 &  02:45:33  &  $-53$:02:04  &   9.1  &  1.00  &  $ -198 \pm 22 $  & $ -336 \pm 37$ &    $ 1.32 \pm 0.15$ &   0.30  & AS0295 \\ 
ACT-CL J0102-4915 &  01:02:53  &  $-49$:15:19  &   9.0  &  0.75  &  $ -520 \pm 58 $  & $ -1046 \pm 116$ &  $ 4.13 \pm 0.46$ &   0.75  & New \\ 
ACT-CL J0438-5419 &  04:38:19  &  $-54$:19:05  &   8.0  &  0.50  &  $ -214 \pm 27 $  & $ -596 \pm 74$ &    $ 2.35 \pm 0.29$ &   0.54  & New \\ 
ACT-CL J0645-5413 &  06:45:30  &  $-54$:13:39  &   7.1  &  0.75  &  $ -221 \pm 31 $  & $ -444 \pm 63$ &    $ 1.75 \pm 0.25$ &   0.17  & A3404 \\ 
ACT-CL J0546-5345 &  05:46:37  &  $-53$:45:32  &   6.5  &  1.75  &  $ -178 \pm 27 $  & $ -240 \pm 37$ &    $ 0.95 \pm 0.15$ &   1.07  & SPT-CL 0547-5345 \\ 
ACT-CL J0235-5121 &  02:35:52  &  $-51$:21:16  &   6.2  &  4.00  &  $ -320 \pm 51 $  & $ -363 \pm 58$ &    $ 1.43 \pm 0.23$ &   0.43  & New \\ 
ACT-CL J0330-5227 &  03:30:54  &  $-52$:28:04  &   6.1  &  1.25  &  $ -145 \pm 24 $  & $ -221 \pm 36$ &    $ 0.85 \pm 0.14$ &   0.44  & A3128(NE) \\ 

\\ \multicolumn{10}{c}{SNR $<$ \highSNR\ and Optically Confirmed} \\ \\
ACT-CL J0616-5227 &  06:16:36  &  $-52$:27:35  &   5.9  &  0.50  &  $ -155 \pm 26 $  & $ -431 \pm 73$   & $ 1.70 \pm 0.29$   &  0.71  & New \\ 
ACT-CL J0559-5249 &  05:59:43  &  $-52$:49:13  &   5.1  &  2.25  &  $ -155 \pm 31 $  & $ -194 \pm 38$   & $ 0.77 \pm 0.15$  &  0.61  & SPT-CL J0559-5249 \\ 
ACT-CL J0215-5212 &  02:15:18  &  $-52$:12:30  &   4.9  &  0.25  &  $ -140 \pm 29 $  & $ -801 \pm 164$  & $ 3.16 \pm 0.65$   &  0.51  & New \\ 
ACT-CL J0509-5341 &  05:09:21  &  $-53$:42:05  &   4.8  &  0.50  &  $ -115 \pm 24 $  & $ -319 \pm 67$   & $ 1.27 \pm 0.27$   &  0.46  & SPT-CL 0509-5342 \\ 
ACT-CL J0641-4949 &  06:41:35  &  $-49$:48:32  &   4.7  &  0.25  &  $ -283 \pm 60 $  & $ -1625 \pm 375$ & $ 6.41 \pm 1.48$   &  0.15  & A3402 \\ 
ACT-CL J0232-5257 &  02:32:45  &  $-52$:57:08  &   4.7  &  0.25  &  $ -124 \pm 26 $  & $ -710 \pm 152$  & $ 2.80 \pm 0.60$   &  0.59  & New \\ 
ACT-CL J0516-5430 &  05:16:30  &  $-54$:30:30  &   4.6  &  1.75  &  $ -143 \pm 31 $  & $ -192 \pm 43$   & $ 0.77 \pm 0.17$  &  0.29  & AS0520/SPT-CL 0517-5430 \\ 
ACT-CL J0346-5438 &  03:46:51  &  $-54$:38:54  &   4.4  &  1.25  &  $ -131 \pm 30 $  & $ -198 \pm 45$   & $ 0.78 \pm 0.18$  &  0.55  & New \\ 
ACT-CL J0217-5245 &  02:17:11  &  $-52$:45:20  &   4.1  &  1.25  &  $ -98 \pm 24 $  & $ -150 \pm 37$    & $ 0.59 \pm 0.14$  &  0.34  & RXC J0217.2-5244 \\ 
ACT-CL J0145-5301 &  01:44:59  &  $-53$:01:01  &   4.0  &  1.00  &  $ -99 \pm 25 $  & $ -167 \pm 42$    & $ 0.65 \pm 0.16$   &  0.12  & A2941 \\ 
ACT-CL J0237-4939 &  02:37:03  &  $-49$:39:27  &   3.9  &  1.00  &  $ -140 \pm 36 $  & $ -236 \pm 61$   & $ 0.93 \pm 0.24$   &  0.40  & New \\ 
ACT-CL J0304-4921 &  03:04:15  &  $-49$:21:42  &   3.9  &  1.00  &  $ -169 \pm 43 $  & $ -285 \pm 74$   & $ 1.12 \pm 0.29$   &  0.47  & New \\ 
ACT-CL J0707-5522 &  07:07:13  &  $-55$:22:48  &   3.3  &  1.75  &  $ -218 \pm 67 $  & $ -293 \pm 90$   & $ 1.16 \pm 0.35$  &  0.43  & New \\ 
ACT-CL J0528-5259 &  05:28:03  &  $-52$:59:53  &   3.1  &  0.50  &  $ -88 \pm 28 $  & $ -244 \pm 78$    & $ 0.94 \pm 0.30$   &  0.77  & SPT-CL 0528-5300 \\ 

\enddata
\tablenotetext{a}{The error simply reflects the SNR with no systematic uncertainty included. }
\tablenotetext{b}{Temperature decrement value with the effect of the ACT beam deconvolved $\delta T_0\times C^{-1}(\theta_{\rm c})$.}
\tablenotetext{c}{Provided for reference: see \citet{menanteau/etal:prep} for the sources of redshifts.}
\label{tab:clusters}
\end{deluxetable}
\clearpage
\end{landscape}

\begin{deluxetable}{cr}
\tabletypesize{\footnotesize}
\tablecolumns{2} 
\tablecaption{Cluster Mass Estimates }
\tablehead{
           \colhead{ACT ID\tablenotemark{a}}                                    &
           \colhead{$M_{500c}$\tablenotemark{b}}  \\
           \colhead{} &
           \colhead{($\times 10^{14} M_\odot$)}
           }
\startdata
ACT-CL J0102-4915 & 9.9 $\pm$ 3.5 \\
ACT-CL J0235-5121 & 6.2 $\pm$ 1.4 \\
ACT-CL J0245-5302 & 6.6 $\pm$ 1.0 \\
ACT-CL J0330-5227 & 9.7 $\pm$ 1.7 \\
ACT-CL J0438-5419 & 12.1 $\pm$ 2.2 \\
ACT-CL J0546-5345 & 5.2 $\pm$ 2.2 \\
ACT-CL J0559-5249 & 5.3 $\pm$ 1.5 \\
ACT-CL J0616-5227 & 6.5 $\pm$ 1.6 \\
ACT-CL J0638-5358 & 9.1 $\pm$ 0.6 \\
ACT-CL J0645-5413 & 7.6 $\pm$ 0.6 \\
ACT-CL J0658-5557 & 14.0 $\pm$ 1.2 \\

\enddata
\tablenotetext{a}{This is a subset of clusters from Table \ref{tab:clusters} with SNR~$>$~5.}
\tablenotetext{b}{Masses derived from X-ray luminosities using the $M-L_X$ relation in \cite{vikhlinin/etal:2009}. The error only includes statistical error propagated from error on $L_X$. }
\label{tab:act_cluster_masses}
\end{deluxetable}

\appendix

\section{Self-similar Scaling Relation for $y_0-M$}
\label{app:scaling}

Assuming an isothermal ICM and taking $R$ as the radius of the cluster, the central Compton $y$ parameter is proportional to the product of the electron temperature ($T$), average central electron number density ($n_0$), and $R$:

\begin{equation}
y_0 \propto  n_{0} T R.
\label{equ:y_prop}
\end{equation}

The electron density can be related through a mean molecular weight to the gas density which, assuming a standard gas fraction, can in turn be related to the total density ($\rho$) of dark matter plus gas. The total density is referenced to the background critical density ($\rho~\propto~\rho_c$) which evolves with redshift as $E^2(z) = \Omega_M(1+z)^3 + \Omega_\Lambda$ in a flat $\Lambda$CDM universe. Furthermore, the cluster radius is a function of the mass and density such that

\begin{equation}
 R \propto \left(\frac{M}{\rho}\right)^{1/3} \propto  \left(\frac{M}{E^2(z)}\right)^{1/3}.
 \label{equ:R_prop}
\end{equation}

Under the assumptions of virialization and hydrostatic equilibrium, the temperature can be related to $M$ and $E(z)$ \cite[e.g.,][]{bryan/norman:1997}:

\begin{equation}
T \propto M^{2/3} E^{2/3}(z).
\label{equ:T_prop}
\end{equation}

Combining these relations, the central Compton $y$ parameter is seen to scale with mass and redshift as

\begin{equation}
y_0 \propto ME^2(z).
\end{equation}

\end{document}